\documentclass[fleqn,usenatbib]{mnras}

\usepackage{newtxtext,newtxmath}
\usepackage[T1]{fontenc}
\usepackage{ae,aecompl}
\usepackage{graphicx}	
\usepackage{amsmath}
\usepackage{amssymb}
\usepackage{color}

\title[Microlensed Radio Emission from Exoplanets]{Microlensed Radio Emission from Exoplanets}

\author[Y. Shiohira et al.]{
Yuta Shiohira$^{1}$\thanks{E-mail: 193d8062@st.kumamoto-u.ac.jp},
Yuka Terada$^{2}$,
Den Mukuno$^{1}$,
Yuka Fujii$^{3,4}$
and Keitaro Takahashi$^{1,5}$
\\
$^{1}$ Graduate School of Science and Technology, Kumamoto University, JAPAN \\
$^{2}$Department of Astronomy, Graduate School of Science, The University of Tokyo, JAPAN \\
$^{3}$National Astronomical Observatory of Japan, Tokyo, 181-8588, JAPAN \\
$^{4}$Earth-Life Science Institute, Tokyo Institute of Technology, Tokyo, 152-8550, JAPAN \\
$^{5}$International Research Organization for Advanced Science and Technology, Kumamoto University, JAPAN
}

\date{Accepted XXX. Received YYY; in original form ZZZ}

\pubyear{2020}

\begin{document}
\label{firstpage}
\pagerange{\pageref{firstpage}--\pageref{lastpage}}
\maketitle

\begin{abstract}
In this paper, we investigate the detectability of radio emission from exoplanets, especially hot Jupiters, which are magnified by gravitational microlensing. Because hot Jupiters have orbital periods much shorter than the characteristic timescale of microlensing, the magnification curve has a unique wavy feature depending on the orbital parameters. This feature is useful to identify radio emission from exoplanets and, in addition to magnification, makes it easier to detect exoplanets directly.
We also estimate the expected event rate red of the detectable level of  microlensed planetary radio emissions, assuming the LOFAR and the first phase of the Square Kilometre Array.
\end{abstract}

\begin{keywords}
gravitational lensing: micro -- radio continuum: planetary systems
\end{keywords}

\section{Introduction}

As the number of discovered exoplanets explodes, the follow-up observations to characterize them from various aspects are being rapidly developed. 
While the majority of exoplanet characterization is conducted in the ultraviolet, optical, and infrared domains, observations at the low-frequency radio wavelengths potentially provide a significant complementary approach to study exoplanets.
In particular, planetary auroral radio emission associated with the interaction of the planetary magnetic field with the stellar wind or with the plasmas from the moons has been proposed as a unique tool to constrain the planetary magnetic fields, as the maximum frequency of the auroral radio spectra is proportional to the magnetic field strength just above the planetary surface. 
The maximum frequency of Jupiter's auroral emission (``cut-off'' frequency), $\sim 30$~MHz, corresponds to $\sim 10$~G of the magnetic field near the surface; A planet with a 10 times stronger surface magnetic field would emit auroral radio up to $\sim 300$~MHz. 
Importantly, planetary auroral emission can be stronger than the emission from the host star (as is the case of the Jupiter-Sun system). 
In other words, the planetary emission may dominate the signal from the system so that it is in principle detectable without any specialized instrument such as a coronagraph. 

Despite the improved planet-to-star contrast, the absolute luminosity is not high. 
For example, the radio emission from an Jupiter-twin at 10 parsec would be on the order of $\sim \mu $Jy. 
Therefore, it has been critical to identify the targets that produce as strong emission as possible, in order to optimize the observations. 
\citet{Zarka1997} estimated the power of auroral radio emission from exoplanets based on so-called ``magnetic Bode's law'', the empirical proportionality between the magnetic energy received by planetary magnetosphere and the power of planetary radio emission, as observed in solar system planets \citep{DeschKaiser1984, Zarka1992}. 
A simple extrapolation of the Bode's law implies that close-in planets (i.e., those with the semi-major axis around or less than 0.1~AU) may produce $10^3-10^6$ times stronger auroral emission than Jupiter due to the enhanced magnetic energy of the stellar wind, provided that the planet has a similar magnetic field as Jupiter \citep[e.g.,][]{christensen2009}.
It should be noted, however, that the planetary radio emission may be suppressed due to the planet's extended atmosphere \citep{daley-yates2018}.
Also, planets around the stars with the stronger stellar wind (e.g., young stars or evolved stars) may be strong radio emitters \citep[e.g.,][]{Griessmeier2007,Fujii+2016}.

Partly assisted by these theoretical expectations, a number of attempts to obtain the evidence of auroral radio emission from discovered exoplanets have been made \citep{winglee1986, bastian2000, george2007, smith2009, lazio2010, stroe2012, hallinan2013, o'gorman2018, lynch2018, sirothia2014, murphy2015}. 
So far, no clear detection has been reported, while there are some possible hints \citep{lecavelier2013, Turner2019, vedantham2020}. 
The upgrade to the existing observatories \citep[e.g., GMRT][]{} or the new capabilities \citep[e.g., NenuFAR][]{} are facilitating the more sensitive search toward the first univocal detection. 
Furthermore, the planned Square Kilometer Array (SKA) will conduct a sensitive survey at this frequency domain, and will provide the number of candidates. 

Motivated by these developments toward searching for exoplanetary  auroral emission, we study the exoplanetary auroral emission observed through gravitational microlensing. 
We consider the scenario in which the planetary system with a hot Jupiter is the {\it source} of the lensing event\footnote{%
This is in contrast to the conventional microlensing method to discover exoplanets in which exoplanets are with the {\it lens} stars.}%
and the planetary auroral emission dominates the light from the source system at the considered wavelength/frequency (for microlensing event of radio emission from extra-terrestrial intelligence, see \cite{rahvar2016}).
In general, gravitational microlensing amplifies the light from the source and allows us to detect the signal at a larger distance, typically closer to the galactic center where the occurrence rate of gravitational microlensing is high. 
Therefore, using microlensing, we can in principle study the magnetic properties of exoplanets at locations different from many known exoplanets. 

The condition that the emission predominantly comes from the planet, the lighter body in the system, causes a notable difference from the microlensing magnification of the planetary system in the optical or infrared domains \citep{graff2000, sajadian2010,bagheri2019}. 
Because the source orbits the star with the period of $O(1)$ days, much shorter than the typical timescale of magnification, the magnification curve has a characteristic wavy feature depending on the orbital parameters. 
This would allow us to identify the planetary nature of the emission, and constrain some of the orbital parameters as well as the properties of the radio emission. 
We discuss these characteristic patterns in the magnification curve and evaluate the occurrence probability of such events. 

The organization of this paper is as follows. 
In section \ref{s2}, we introduce the framework of gravitational microlensing of an exoplanet as a source. 
Next, we show magnification curves and consider the relation between their feature and parameters of the system in section \ref{s3}. 
In section \ref{s4}, we discuss the observability of distant hot Jupiters through microlensing. 
Section \ref{s5} is devoted to discussion and summary.

\section{Microlensing of exoplanets}
\label{s2}

\begin{figure*}
\centering
\includegraphics[width=9cm]{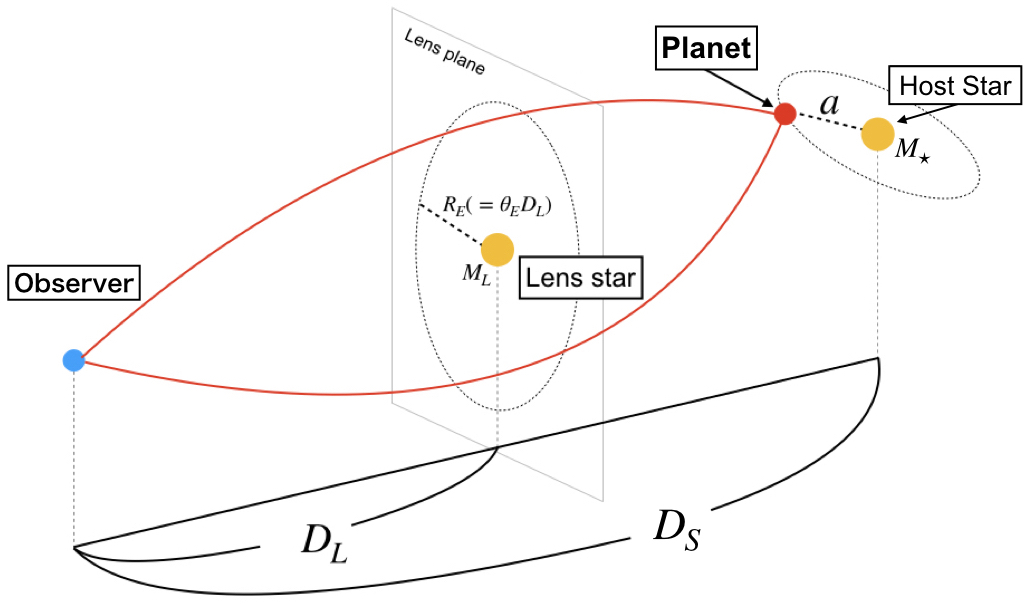}\includegraphics[width=7cm]{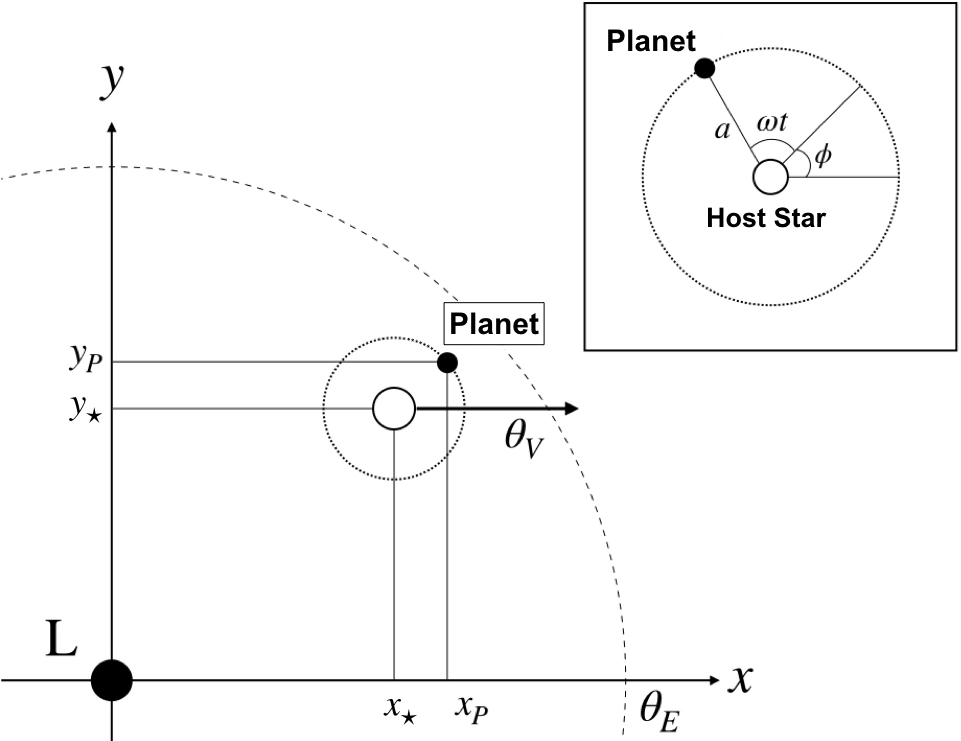}
\caption{Schematic view of the configuration considered in this paper (left) and the coordinates on the lens plane (right). In the left panel, $M_L$ and $M_{\star}$ are the masses of the lens star and host star, respectively, and $D_L$ and $D_S$ are the distance from the observer to the lens star and source star, respectively. In the right panel, the lens star is located at the origin on the lens plane. The position of the host star and planet are designated as $(x_{\star}, y_{\star})$ and $(x_P, y_P)$, respectively. $\theta_V$ is the relative transverse speed of lens and host star on the lens plane, $a$, $\omega$ and $\phi$ are the semi-major axis, angular velocity and initial orbital phase of the planet, respectively.}
\label{fig_ptrj1}
\end{figure*}

Here we present a basic formalism to study microlensing of an exoplanet orbiting around the host star. The configuration we are considering is depicted in Fig. \ref{fig_ptrj1}. First, the Einstein radius, $\theta_E$, is defined as,
\begin{eqnarray}
\label{radi_ein}
\theta_{E}
&\equiv& \sqrt{\frac{4 G M_L}{c^{2}} \frac{D_{S}-D_{L}}{D_{S} D_{L}}} \nonumber \\
&\sim& 0.001~\mathrm{[arcsec]}
       \left(\frac{M_L}{M_{\odot}}\right)^{\frac{1}{2}}
       \left(\frac{7[\mathrm{kpc}]}{D_{S}}\right)^{\frac{1}{2}}
       \left(\frac{D_{S}-D_{L}}{D_{L}}\right)^{\frac{1}{2}}
\quad ,
\end{eqnarray}
where $M_L$ is the lens mass, $D_L$ and $D_S$ are the distance from the observer to the lens and source, respectively. When the host star passes through the Einstein ring, the rate at which the angular distance between the lens and the host star changes is given by
\begin{eqnarray}
\label{velocity}
\dot \theta 
&=& \frac{V}{D_{L}} \nonumber \\
&=& 0.012~\mathrm{[arcsec/}\mathrm{yr]}
    \left(\frac{V}{200\left[\mathrm{km/s}\right]}\right)
    \biggl(\frac{3.5\mathrm{[kpc]}}{D_{L}}\biggl)
\end{eqnarray}
where $V$ is the relative transverse speed of the lens and host star on the lens plane. The crossing time, a characteristic timescale for the host star to cross the Einstein ring, $t_E$, is given as,
\begin{eqnarray}
\label{time_ein}
t_{E}
&=& \frac{\theta_{E}}{\dot \theta}\nonumber \\
&\sim& 0.26~\mathrm{[yr]}
       \left(\frac{M_L}{M_{\odot}}\right)^{\frac{1}{2}} 
       \left(\frac{D_{L}}{3.5~[\mathrm{kpc}]}\right)^{\frac{1}{2}}
       \left(\frac{D_{S}-D_{L}}{D_{S}}\right)^{\frac{1}{2}}
\nonumber \\
& &    \times
       \left(\frac{100~\left[\mathrm{km/s}\right]}{V}\right).
\end{eqnarray}
The magnification due to microlensing, $A$, is given by,
\begin{eqnarray}
\label{amp}
A = \frac{2 + u^{2}}{u \sqrt{u^{2} + 4}}
\end{eqnarray}
where $u = \theta_S / \theta_E$ and $\theta_S$ is the angular distance between the lens and source (planet). We denote the minimum angular distance between the lens and host star (rather than the planet) divided by the Einstein radius as $u_{\mathrm{min}}$.

As shown in the right panel of Fig. \ref{fig_ptrj1}, we assume that the host star moves in the $x$ direction on the lens plane and the motion is expressed as $(x_{\star},y_{\star}) = (\theta_V t,y_{\star})$. Then, the motion of the planet (source) around the host star projected on the lens plane is expressed as,
\begin{eqnarray}
\left( \begin{array}{cc} x \\ y \\ \end{array} \right)
=
\left( \begin{array}{cc}
x_{\star} \\ y_{\star} \\
\end{array} \right)
+
\frac{a}{D_S}
\left( \begin{array}{cc} 
\cos{\Omega} & -\sin{\Omega} \cos{i} \\
\sin{\Omega} & \cos{\Omega} \cos{i} \\
\end{array} \right)
\left( \begin{array}{cc}
\cos{(\omega t + \phi)} \\ \sin{(\omega t + \phi)} \\
\end{array} \right), \nonumber \\
\label{peq}
\end{eqnarray}
where $a$ is the semi-major axis, $\omega$ is the angular velocity of orbital motion, $\phi$ is the initial orbital phase of the planet, $i$ is the orbital inclination and $\Omega$ is longitude of the ascending node. Here we assume a circular orbit for simplicity. Thus, the current system is characterized by 10 parameters: $M_{\star}$, $M_L$, $D_S$, $D_L$, $a$, $u_{\mathrm{min}}$, $V$, $\phi$, $i$ and $\Omega$. By substituting Eqs. (\ref{peq}) into Eq. (\ref{amp}), we can calculate the evolution of magnification of the planet. Finally, it is instructive to see the angular scale of the semi-major axis divided by the Einstein radius,
\begin{eqnarray}
u_\mathrm{a}
&\equiv& \frac{a/D_S}{\theta_E}
\nonumber \\
&=&
1.5 \times 10^{-2}
\left( \frac{a}{0.1~\mathrm{AU}} \right)
\left( \frac{M_L}{M_\odot} \right)^{-\frac{1}{2}}
\left( \frac{D_S}{7~\mathrm{kpc}} \right)^{-\frac{1}{2}}
\left( \frac{D_L}{D_S - D_L} \right)^{\frac{1}{2}}.
\nonumber \\
\label{ua}
\end{eqnarray}

\section{Magnification curve}
\label{s3}

In this section, we show the magnification curves of microlensed exoplanets calculated from the formulation given in the previous section. Here, it should be noted that we are considering a case where the planet is intrinsically much brighter than the host star so that only the planet can be observed even in the presence of microlensing if the magnifications are comparable.

\begin{table}
    \centering
        \label{p}
        \begin{tabular}{cc}
            \hline
    		    Parameters & Values\\
    		\hline
            	$M_{\star}$& $M_{\star}=1.0M_{\odot}$\\
            	$M_L$&$M_L=0.1M_{\odot},1.0M_{\odot}$\\
        		$D_S$&\\
        		$D_L$&\\
        	    $a$ & $R_{\odot} \leq a \leq 0.1\mathrm{AU}$\\
        	    $u_{\mathrm{min}}$ & $ 0 \leq u_{\mathrm{min}} \leq 1$\\
        	    $V$&\\
        	    $\phi$&$0^{\circ}\leq \phi \leq 360^{\circ}$\\
        		$i$&$ 0^{\circ}\leq i \leq 180^{\circ}$\\
        		$\Omega$&$0^{\circ}\leq \Omega \leq 180^{\circ}$\\
            \hline
        \end{tabular}
        \caption{Parameters and their values.}
\end{table}

\begin{figure*}
\centering
\includegraphics[width=6.2cm]{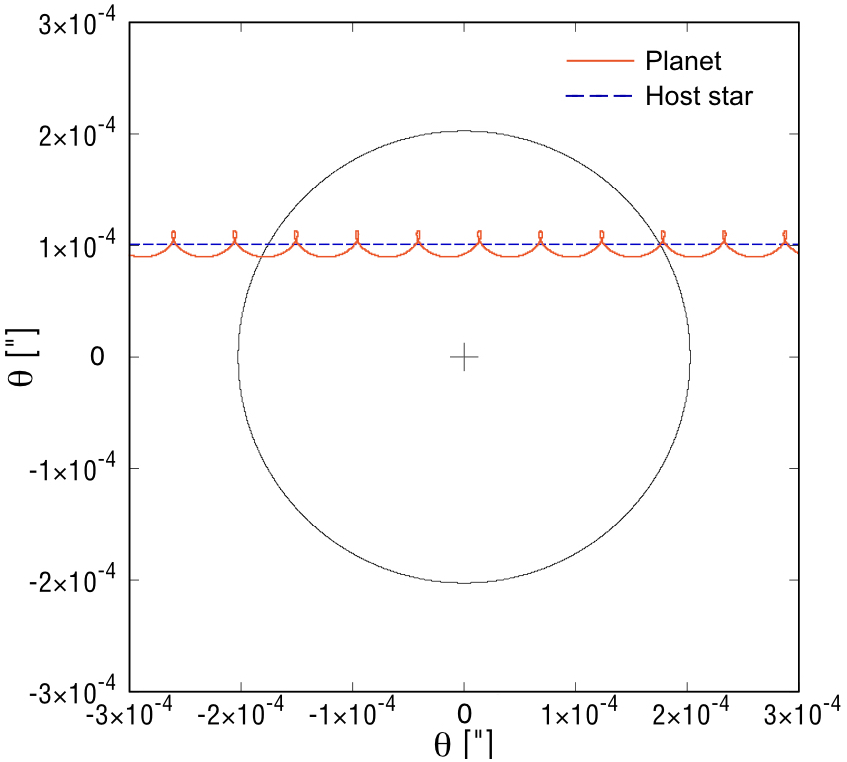}
\quad\quad
\includegraphics[width=6cm]{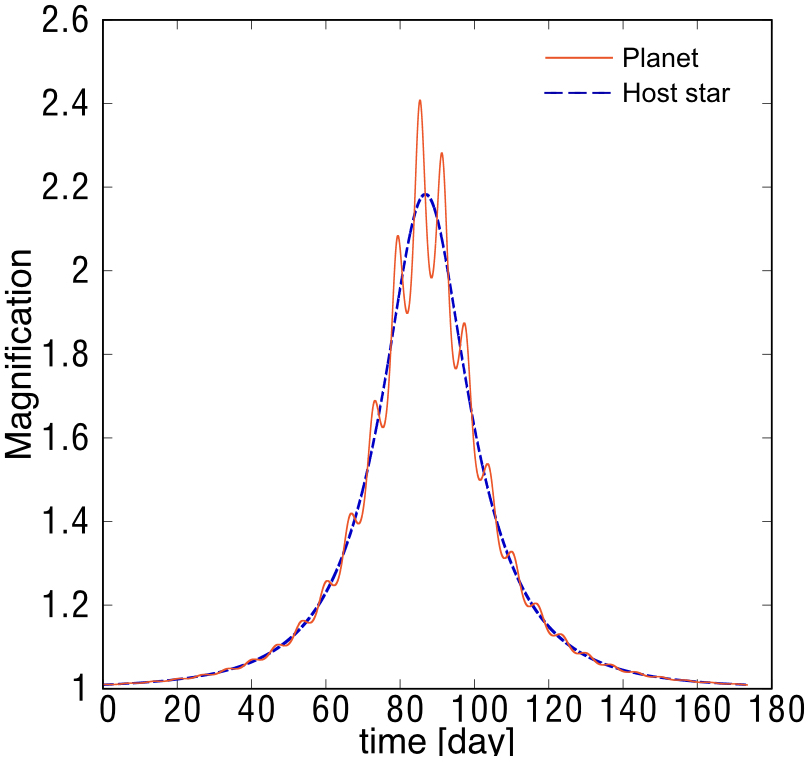}
\caption{Left: trajectories of the planet (solid) and host star (dashed) with parameters, $M_{\star} = 3.0~M_{\odot}$, $M_L = 0.2~M_{\odot}$, $D_S = 8.5~\mathrm{kpc}$, $D_L = 7.0~\mathrm{kpc}$, $a=0.1~\mathrm{AU}$, $u_{\mathrm{min}}=0.5$, $V = 100~\mathrm{km/s}$, $\phi = 0^{\circ}$, $i = 0^{\circ}$ and $\Omega=0^{\circ}$. The lens is located at the origin and the circle represents the Einstein radius. Right: corresponding magnification curves.}
\label{fig_p_ex}
\end{figure*}

Fig.~\ref{fig_p_ex} shows an example trajectory and magnification curve of a microlensing event of a planet with a small semi-major axis, $a = 0.1~\mathrm{AU}$. Here, other parameters are set as $M_{\star} = 3.0~M_{\odot}$, $M_L = 0.2~M_{\odot}$, $D_S = 8.5~\mathrm{kpc}$, $D_L = 7.0~\mathrm{kpc}$, $u_{\mathrm{min}}=0.5$, $V = 100~\mathrm{km/s}$, $\phi = 0^{\circ}$, $i = 0^{\circ}$ and $\Omega=0^{\circ}$. For comparison, the trajectory and magnification curve of the host star are also shown. The amplitude of the oscillation of the planet's trajectory in $y$ direction is determined by Eq.~(\ref{ua}). As we see in the right plot, the magnification of the planet waves around that of the host star due to the periodic orbital motion. The maximum magnification of the planet exceeds that of the host star because, as seen in the left panel, the planet approaches the lens more than the host star. In fact, the shape of magnification curve and the maximum magnification depends on the initial orbital phase. Below, we investigate the dependence of the magnification curve on the parameters.

\begin{figure*}
\centering
\includegraphics[width=6.2cm]{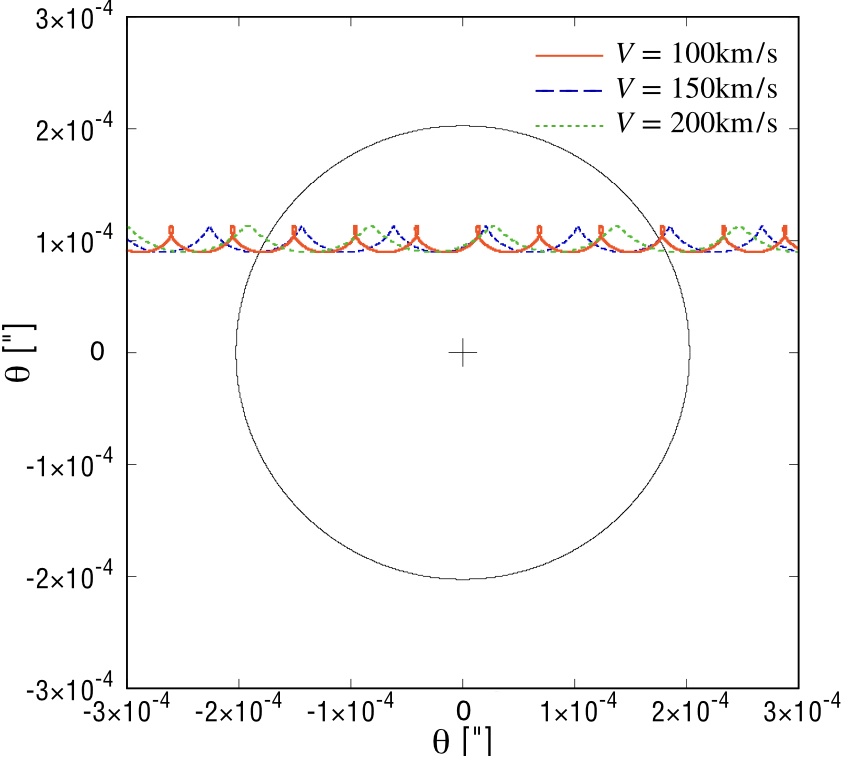}
\quad\quad
\includegraphics[width=6.2cm]{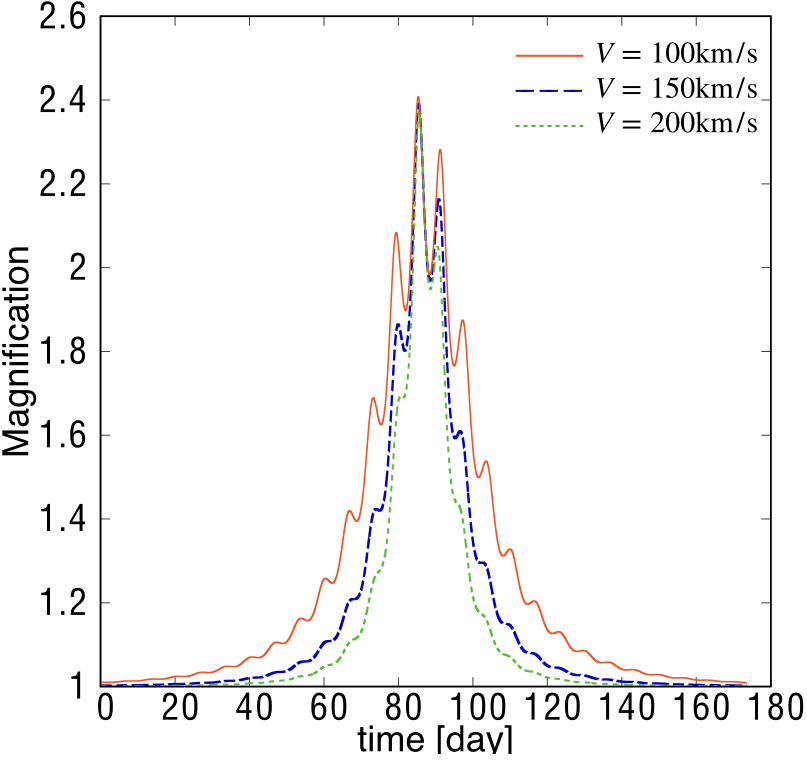}
\caption{Same as Fig.~\ref{fig_p_ex} but varying the relative transverse velocity, $V$. The red solid line, blue dashed line and green dotted line cofrespond to $V = 100~\mathrm{km/s}$, $V = 150~\mathrm{km/s}$ and $V = 200~\mathrm{km/s}$, respectively.}
\label{v}
\end{figure*}

In Fig.~\ref{v}, the variation of the magnification is shown by varying the relative transverse velocity, $V$, and fixing other parameters. The crossing time shortens for a smaller value of $V$ as in Eq.~(\ref{time_ein}) and the magnification curve shrinks in time direction. Furthermore, the wavy feature in the magnification curve of the tail (magnification $\lesssim 1.5$) is reduced when the transverse velocity becomes comparable or larger than the planet's orbital velocity. In this plot the maximum magnifications are almost the same for the three cases but this depends on the initial phases of the orbital motion again.

\begin{figure*}
\centering
\includegraphics[width=6.2cm]{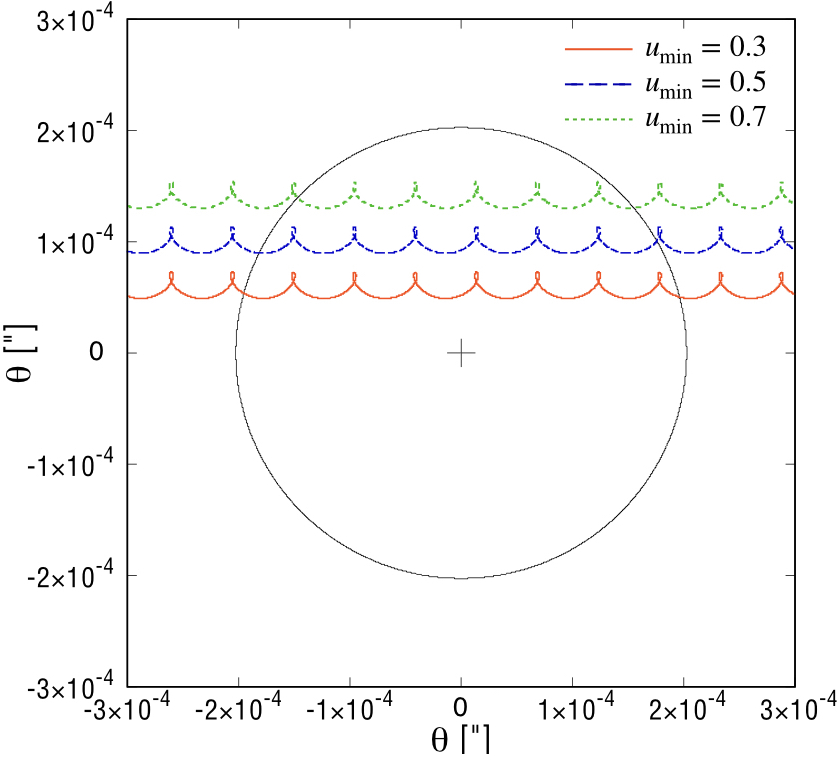}
\quad\quad
\includegraphics[width=6.2cm]{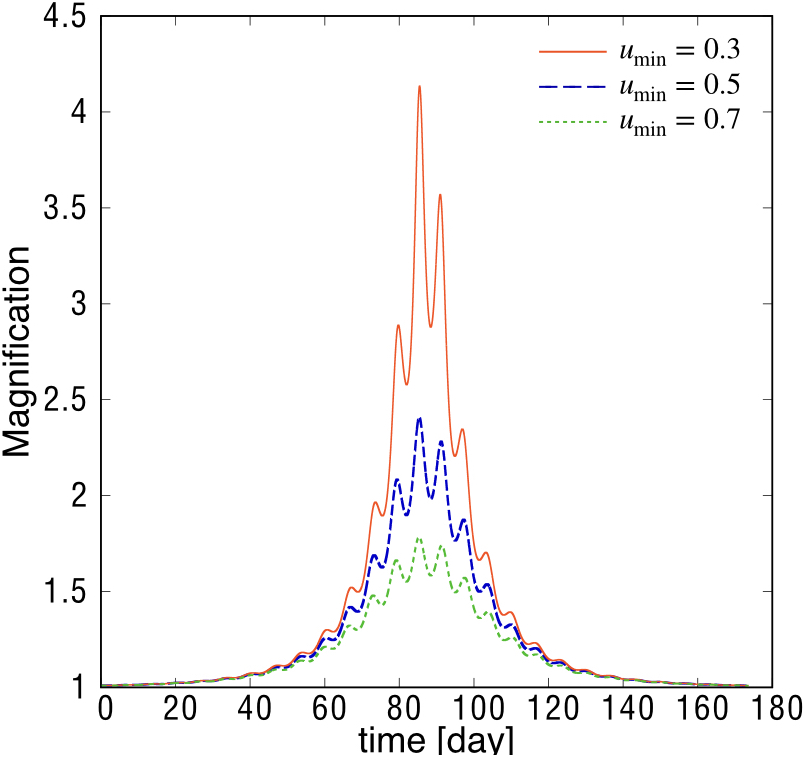}
\caption{Same as Fig.~\ref{fig_p_ex} but varying $u_\mathrm{min}$. The red solid line, blue dashed line and green dotted line correspond to $u_\mathrm{min} = 0.3$, $u_\mathrm{min} = 0.5$ and $u_\mathrm{min} = 0.7$, respectively.}
\label{umin}
\end{figure*}

Fig.~\ref{umin} compares the trajectory and magnification curve for different values of $u_\mathrm{min}$. It is seen that a smaller value of $u_\mathrm{min}$ results in not only overall enhancement of the magnification but also a larger amplitude of the oscillation of the magnification around that of the host star (not shown). This is because the fractional variation of $u$ of the planet becomes larger for a smaller value of $u_\mathrm{min}$.

\begin{figure*}
\centering
\includegraphics[width=6.2cm]{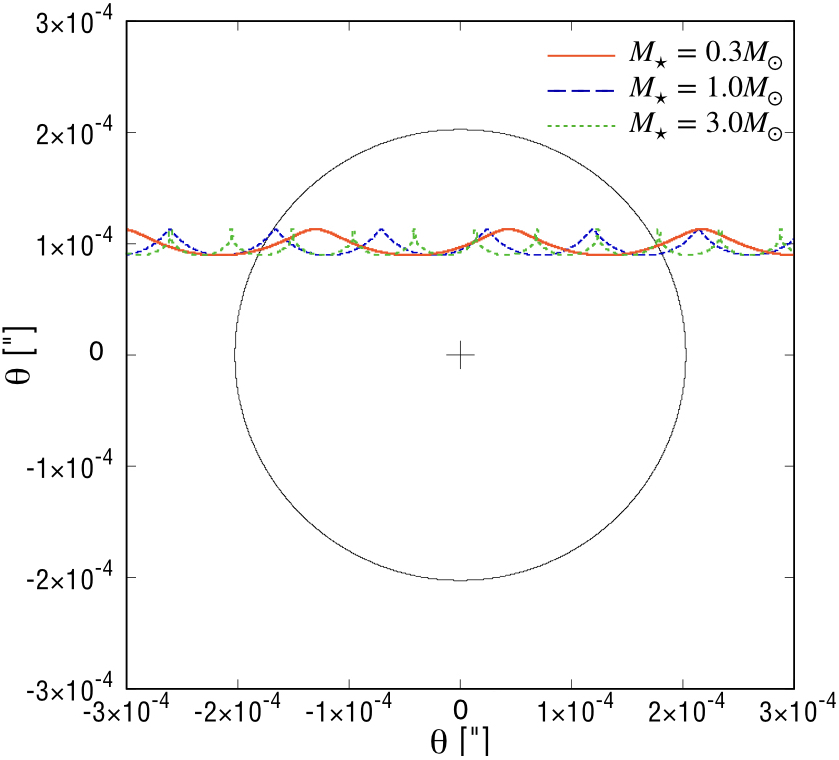}
\quad\quad
\includegraphics[width=6.2cm]{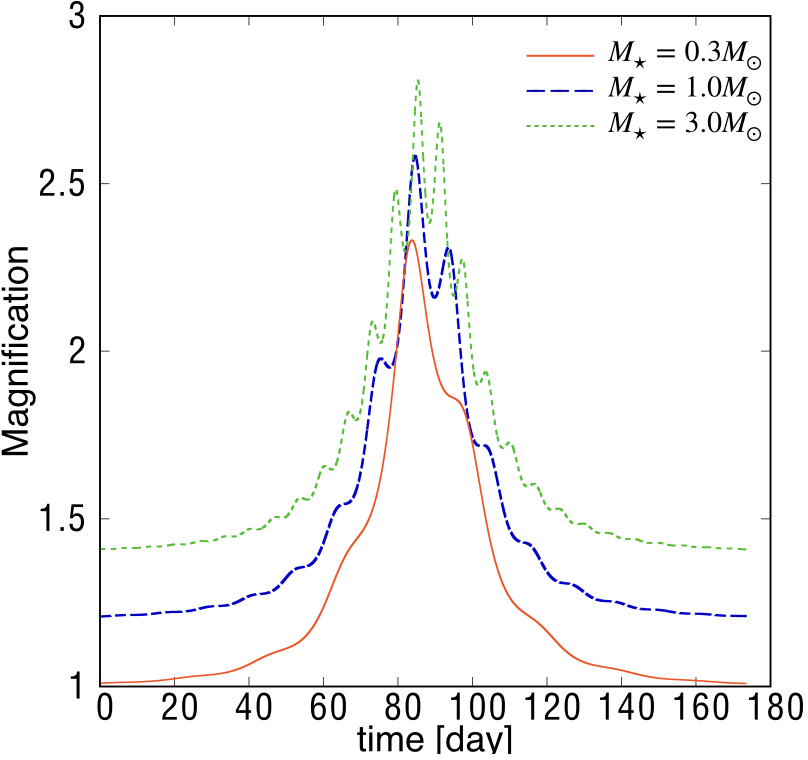}
\caption{Same as Fig.~\ref{fig_p_ex} but varying the host star mass $M_\star$. The red solid line, blue dashed line and green dotted line correspond to $M_{\star} = 0.3~M_{\odot}$, $M_{\star} = 1.0~M_{\odot}$ and $M_{\star} = 3.0~M_{\odot}$, respectively. In the right panel, the blue dashed line and green dotted line are shifted upward by 0.2 and 0.4, respectively.}
\label{ms}
\end{figure*}

\begin{figure*}
\centering
\includegraphics[width=6.2cm]{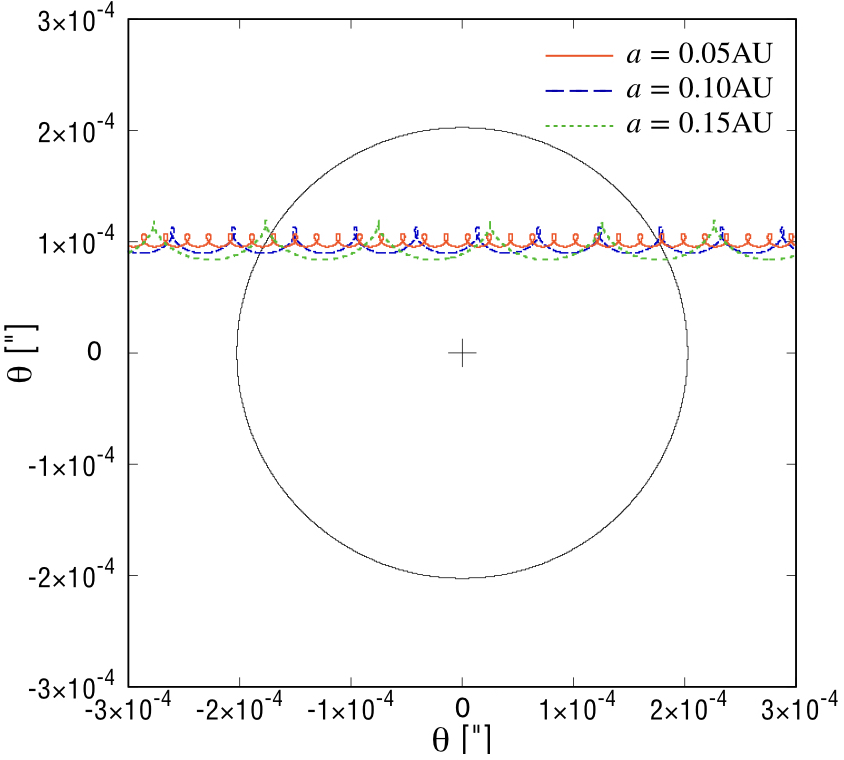}
\quad\quad
\includegraphics[width=6cm]{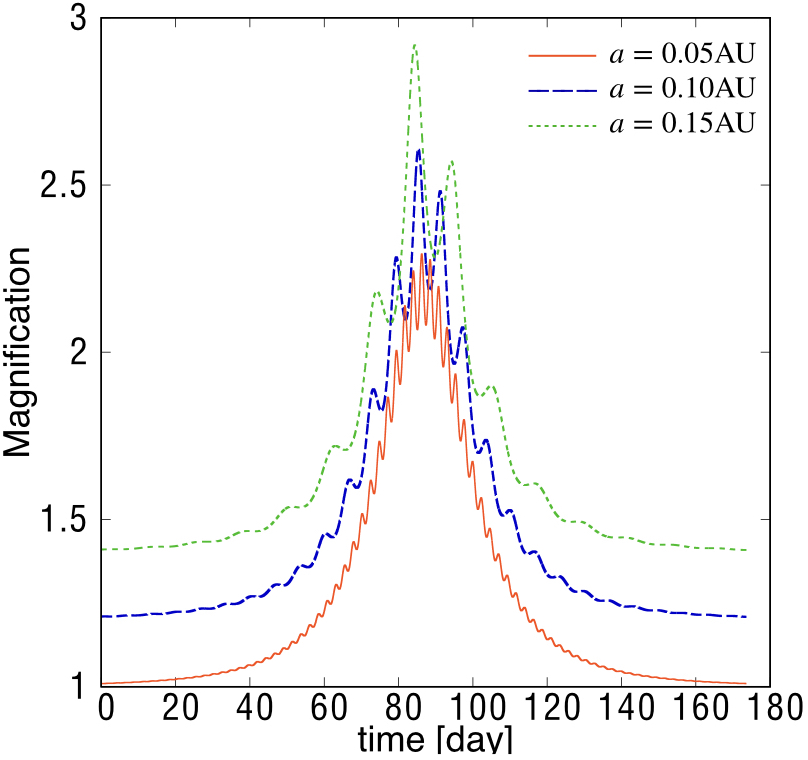}
\caption{Same as Fig.~\ref{fig_p_ex} but varying the semi-major axis $a$. The red solid line, blue dashed line and green dotted line correspond to $a = 0.05~\mathrm{AU}$, $a = 0.10~\mathrm{AU}$ and $a = 0.15~\mathrm{AU}$, respectively. In the right panel, the blue dashed line and green dotted line are shifted upward by 0.2 and 0.4, respectively.}
\label{a}
\end{figure*}

The effect of changing the host star mass is shown in Fig.~\ref{ms}. A smaller mass leads to a longer orbital period and the number of the oscillation of the magnification is reduced for a fixed crossing time (relative transverse velocity). Similar effect can be seen in Fig.~\ref{a}, where the semi-major axis is varied, because the orbital period depends on the semi-major axis as well. In this case, the $y$ range of the planet trajectory is expanded for a larger value of $a$, and the maximum magnification is larger.

\begin{figure*}
\centering
\includegraphics[width=6.2cm]{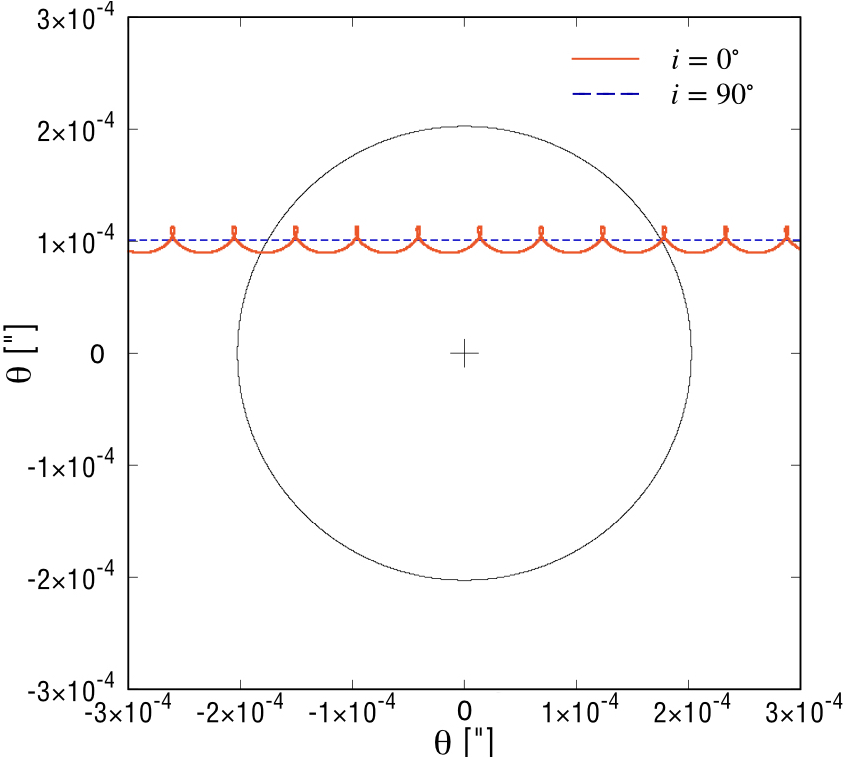}
\quad\quad
\includegraphics[width=6cm]{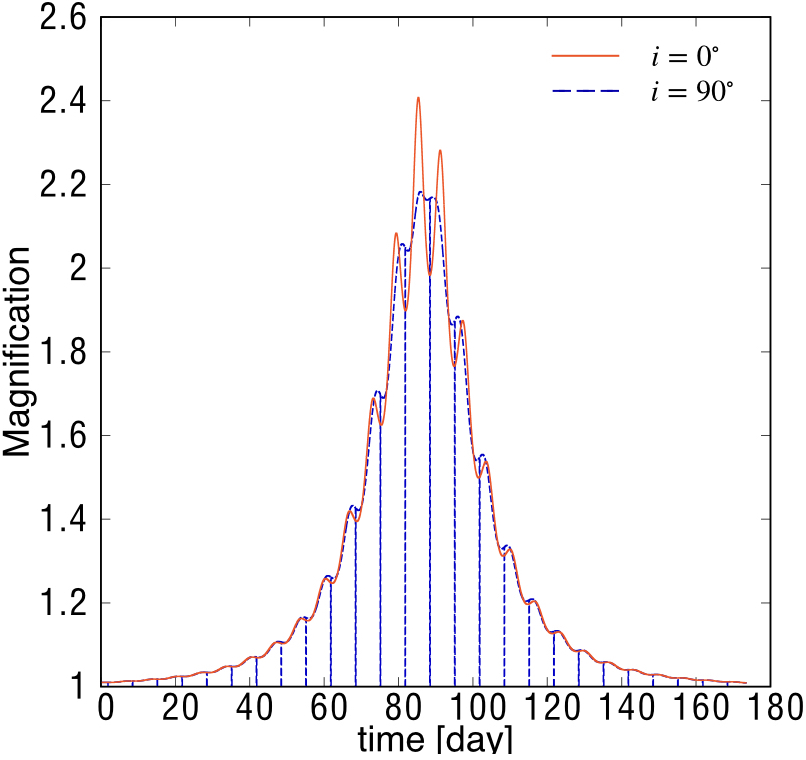}
\caption{Same as Fig.~\ref{fig_p_ex} but varying the inclination $i$. The red solid line and blue dashed line correspond to $i = 0^{\circ}$ (face-on) and $i = 90^{\circ}$ (edge-on), respectively. In the right panel, the vertical lines represent the transit of the planet by the host star.}
\label{i}
\end{figure*}

Finally, Fig.~\ref{i} compares the cases of face-on ($i = 0^{\circ}$) and edge-on ($i = 90^{\circ}$). In the edge-on case, the trajectory is a straight line, but the $x$-coordinate does not increase monotonically and the planet can move backwards in the lens plane. This is why the magnification oscillates even for the edge-on case, although the wavy feature is smeared substantially. The vertical lines represent the transit of the planet by the host star and the radio emission is diminished there.

As we have seen, a source planet with a relatively small semi-major axis, and then a short orbital period, can have a unique wavy feature in the magnification curve, depending on the parameters. The feature is useful for not only the identification of microlensing event of exoplanets but estimation of the fundamental parameters of the planetary system. However, these wavy features cannot be seen when the orbital period is comparable to or longer than the crossing time. In this case, the magnification curve would be very similar to normal microlensing events. Nevertheless, microlensing will be very useful to observe faint radio signals from distant exoplanets.

\section{Event rate}
\label{s4}

In this section, we investigate the observability of the radio emission from exoplanets amplified by gravitational microlensing and estimate the event rate.
We consider the microlensing events toward the Galactic Center because the event rate is expected to be relatively high in the direction.
In fact, the SKA is planning surveys of the Galactic Center and Galactic plane with a survey time of order $O(1000)$ hours.
They are commensal surveys so that the data can be used for various scientific purposes.
We assume to use the data of such large commensal surveys to search for microlensing events.
However, the microlensed sources in the direction of Galactic Center are mostly located at the bulge region with a typical distance of $7~{\rm kpc}$.
Thus, the radio emission is very faint and, as we will see, a magnification of an order of O(100) is required to be detected even with the SKA.

We can estimate the event rate of microlensing of hot Jupiters by multiplying the following factors.
\begin{enumerate}
\item $S~[\mathrm{deg}^2]$: area of survey field
\item $R_\mathrm{lens}~[\mathrm{/year /deg}^2]$: event rate of microlensing of stars
\item $P_\mathrm{HJ}$: probability that a star has a hot Jupiter
\item $P_\mathrm{obs}$: conditional probability that the peak luminosity of a hot Jupiter reaches the detection threshold given that the host star enters the Einstein ring of a lens star
\end{enumerate}
In this study, we assume a survey area of $S = 100~[\mathrm{deg}^2]$.
We estimate $R_{\mathrm{lens}} \approx 100\mathrm{/year /deg^2}$ in the direction of Galactic Center, based on \citet{mroz2019}. 
The occurrence of a hot Jupiter around FGK stars has been estimated to be $P_{\mathrm{HJ}}\sim 1\%$ \citep{mayor2011,wright2012,fressin2013}. 
The remaining number to estimate is $P_\mathrm{obs}$.

Below, we evaluate $P_\mathrm{obs}$ through a Monte-Carlo method considering the variation of the 9 parameters given in the previous section and the emissivity of exoplanets.
To do this, we need to determine the radio power of an exoplanet given a set of parameters.
It is suggested that the radio power is proportional to the incident power of stellar wind into the planet's magnetosphere \citep[e.g.,][]{Zarka1992,Zarka1997}. 
Thus, the radio power may naively be assumed to be proportional to a geometric factor of $a^{-2}$. 
However, as $a$ decreases, the kinetic and magnetic pressure of the stellar wind increases, and the planet's magnetosphere would shrink and the incident power may not increase so rapidly. 
Considering this effect, the radio power would be proportional to $a^{-4/3}$  \citep[e.g.,][]{Griessmeier2005}.
Although there are other scaling laws suggested in the literature \citep{farrell1999, zarka2001, lazio2004}, we consider the above two cases ($a^{-2}$, $a^{-4/3}$) as typical scaling laws. Concerning the overall normalization of radio power, we take Jupiter's one ($P_{\mathrm{rad}} = 10^{11}~\mathrm{W}$ at $a = 5~\mathrm{AU}$). 
Thus, because most of observable exoplanets have a very small semi-major axis ($\lesssim 0.1~\mathrm{AU}$) as we will see later, our targets are mostly hot Jupiters.

In this study, we consider using the SKA1-LOW that will have a high sensitivity of $70~\mu\mathrm{Jy}$ at 50 MHz with $4~\mathrm{MHz} \times 1~\mathrm{hour}$ integration \citep{zarka2015}.
A deeper and wider-band observation ($40~\mathrm{MHz} \times 10~\mathrm{hours}$ integration, designated as "SKA1-LOW Deep") is also considered. As a comparison, we evaluate the potential of observations by the LOFAR with sensitivity which is about 1/30 of the SKA1-LOW at $50~\mathrm{MHz}$. Thus, for example, an exoplanet with $a = 0.05~\mathrm{AU}$ at $1~(7)~\mathrm{kpc}$ needs a magnification of a factor of $\sim 60~(2800)$ to be detectable by SKA1-LOW Deep.

In fact, the highest frequency of Jupiter's auroral radio emission is $\sim 30~\mathrm{MHz}$ and outside the frequency range of the SKA1-LOW. However, the highest frequency is proportional to the magnitude of the surface magnetic field ($12~\mathrm{Gauss}$ for Jupiter) and even stronger magnetic fields are expected for hot Jupiters \citep[e.g.,][]{Cauley2019}. Thus, we assume exoplanet's radio emission extends beyond $50~\mathrm{MHz}$, which is possible if the surface field is stronger than $20~\mathrm{Gauss}$.

\begin{figure*}
\centering
\includegraphics[width=8.0cm]{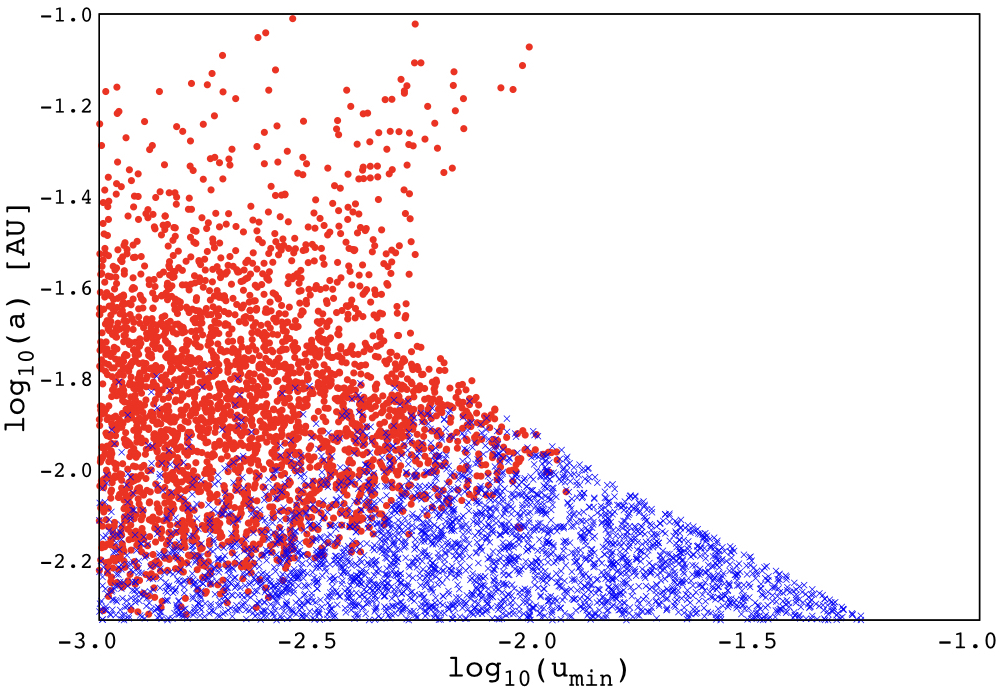}
\includegraphics[width=8.0cm]{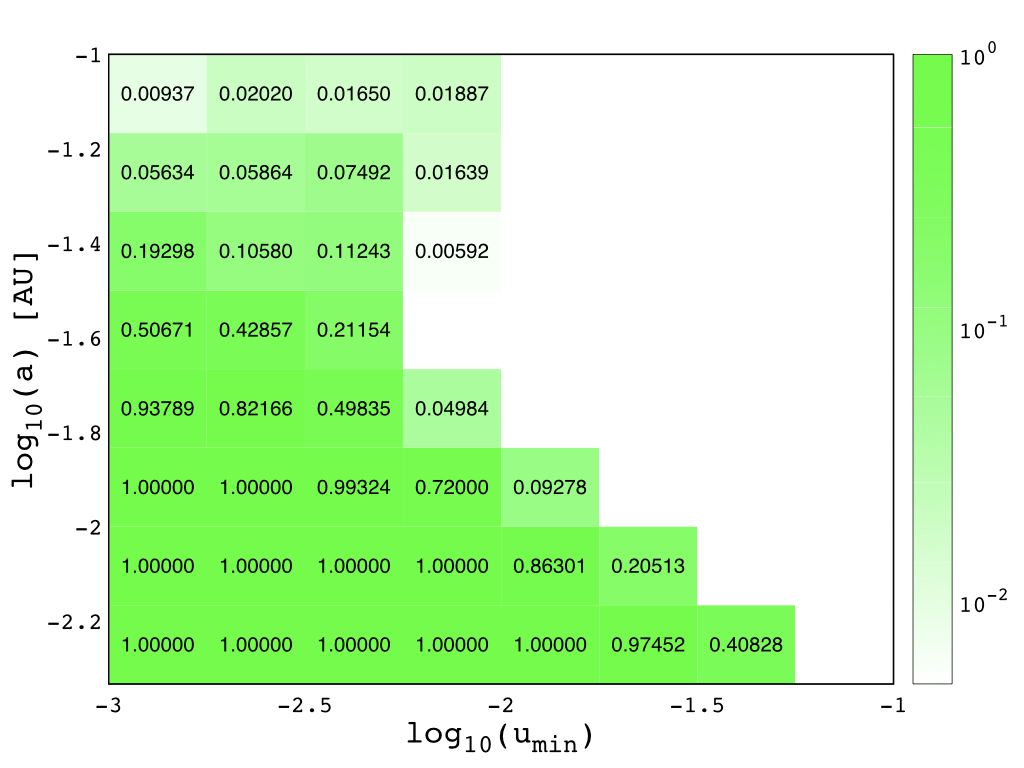}
\caption{Left: scatter plot of exoplanets detectable with SKA1-LOW Deep (40MHz $\times$ 10 hours integration) assuming a scaling law of $a^{-2}$. Four of the parameters, $a$, $u_\mathrm{min}$, $\phi$, $i$ and $\Omega$ are randomly chosen, while other parameters are set as $M_{\star} = 1.0~M_{\odot}$, $M_L = 1.0~M_{\odot}$, $D_S = 7.0~\mathrm{kpc}$, $D_L = 3.5~\mathrm{kpc}$ and $V = 200~\mathrm{km/s}$. Blue and red points represent planets whose magnification curves have multiple peaks and a single peak, respectively. Right: probability of the detection for each region in $a$-$u_\mathrm{min}$ plane.
}
\label{sca}
\end{figure*}

Based on the above assumptions, we can study the detectability of radio emission from exoplanets. 
First, we discuss the dependence on the detectability in the $a$-$u_\mathrm{min}$ plane (Fig. \ref{sca}).
For this, we make 19997 parameter sets by randomly choosing $a$, $u_\mathrm{min}$ (random in the logarithmic scale), $\phi$, $i$ and $\Omega$ (random in the linear scale) while fixing other parameters such that $M_{\star} = 1.0~M_{\odot}$, $M_L = 1.0~M_{\odot}$, $D_S = 7.0~\mathrm{kpc}$, $D_L = 3.5~\mathrm{kpc}$ and $V = 200~\mathrm{km/s}$. Then, for each parameter set, we ask if (i) the amplified radio emission is above the sensitivity of SKA1-LOW Deep, and (ii) if the magnification curve has multiple peaks and the characteristic wavy features, which would be needed to identify the planetary nature of the emission.
For the scaling of radio emissivity, we adopt the $a^{-2}$ law.
The left panel of Fig. \ref{sca} represents the scatter plot where the blue crosses correspond to the parameter sets that satisfy both criteria with SKA1-LOW Deep, i.e., they are detectable by SKA1-LOW Deep with wavy features. On the other hand, red points represents the parameter sets that satisfy the first criterion only, i.e., the radio power is detectable but the wavy patterns are not seen.
The typical value of $u_{\rm min}$ is $10^{-2}$, which corresponds to a magnification factor of $\sim 1/u_{\rm min} \sim 100$.

Generally, in order for an exoplanet to be detectable, values of $a$ and/or $u_\mathrm{min}$ must are smaller than figures described in section 2 because they give a large intrinsic luminosity and large magnification, respectively. 
If the orbital motion of a planet is ignored, the peak luminosity scales as $a^{-2} u_\mathrm{min}$. Therefore, planets under a line $a^{-2} u_\mathrm{min} = \mathrm{const.}$, which goes from up-left to bottom-right in the left panel of Fig. \ref{sca}, are detectable for most of other random parameter sets ($\phi$, $i$ and $\Omega$) (see the right panel). In this case, as we can see, the magnification curve often has multiple peaks.

On the other hand, we see that many planets just above the line are still detectable with the probability of $O(10) \%$. Interestingly, at $u_\mathrm{min} \approx 10^{-2}$, there are a few detectable planets with $a > 0.05~\mathrm{AU}$, while no planets with $0.05~\mathrm{AU} > a > 0.01~\mathrm{AU}$ are found. This is apparently strange because the intrinsic luminosity is larger for planets with $0.05~\mathrm{AU} > a > 0.01~\mathrm{AU}$. In fact, for $a > 0.05~\mathrm{AU}$, the angular scale of the semi-major axis is as large as $u_\mathrm{min}$ (see Eq.~(\ref{ua})) so that the planet can be very close to the lens star depending on the initial orbital phase. Thus, planets with $a > 0.05~\mathrm{AU}$ can have very large magnification of $O(1000)$ and become detectable even if the intrinsic luminosity is relatively low.

\begin{figure*}
\centering
\includegraphics[width=7.0cm]{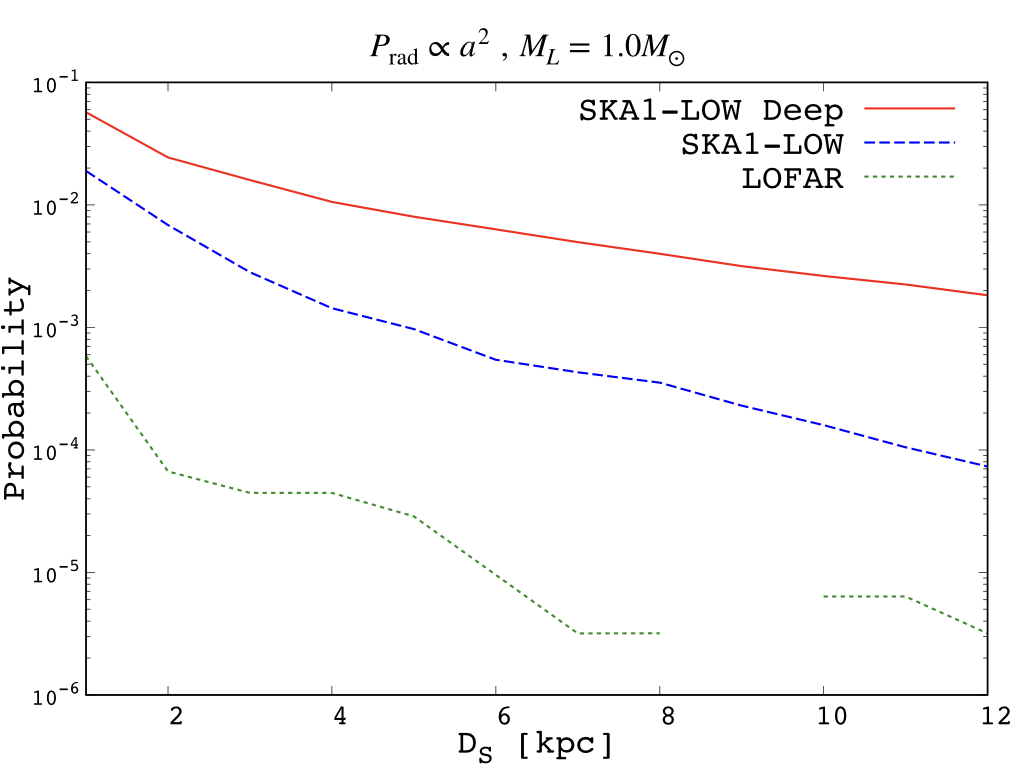}
\quad\quad
\includegraphics[width=7.0cm]{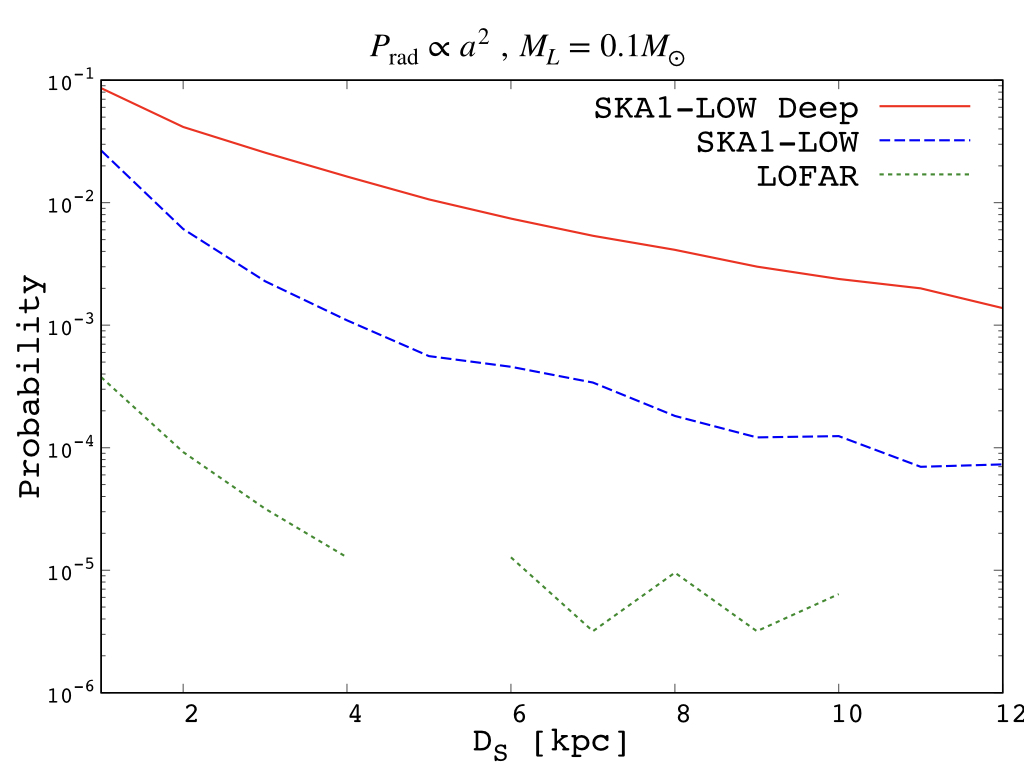} \includegraphics[width=7.0cm]{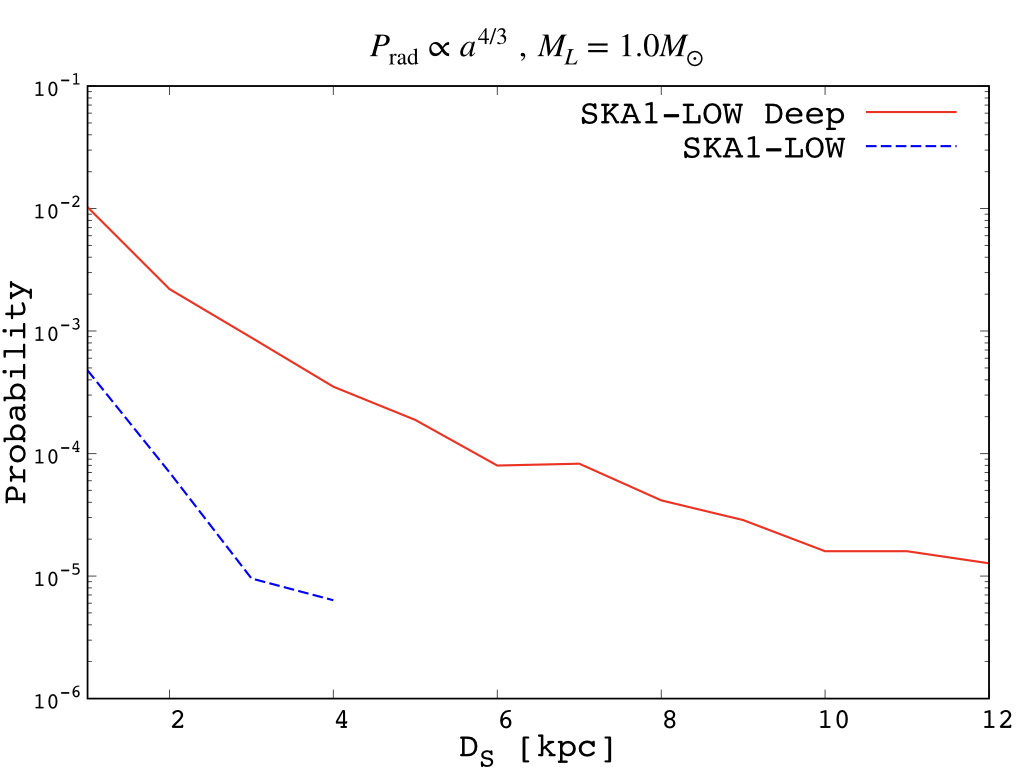}
\quad\quad
\includegraphics[width=7.0cm]{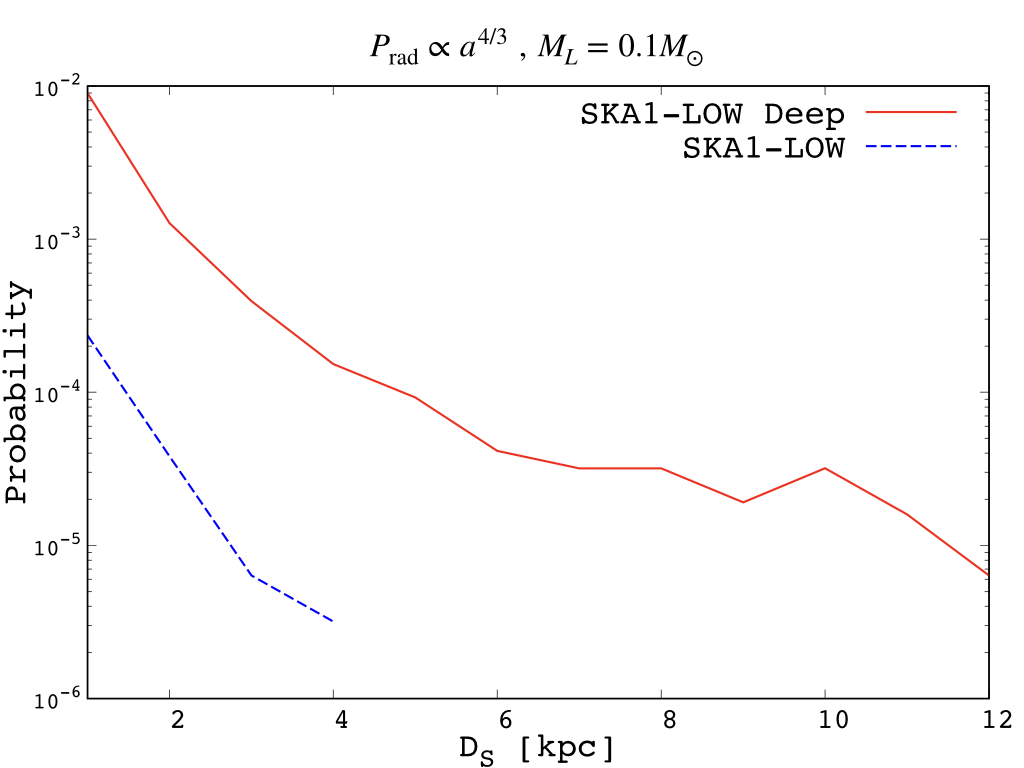}
\caption{Conditional probability that an exoplanet can be detected as a function of the distance, $D_S$, given that the host star enters the Einstein radius of a lens star. In upper and lower panels, the scaling law of $P_\mathrm{rad} \propto a^{-2}$ and $a^{-4/3}$ are assumed, respectively. In left and right panels, the lens mass is set to $0.1~M_{\odot}$ and $1.0~M_{\odot}$, respectively. Here, the lens distance is set to the half of the source distance. The semi-major axis is chosen randomly according to a probability distribution function derived from the observed population [reference]. The values of $u_\mathrm{min}$, $\phi$, $i$ and $\Omega$ are randomly chosen from uniform probability distributions. Other parameters are set as $M_{\star} = 1.0~M_{\odot}$ and $V = 200~\mathrm{km/s}$. Red, blue and green curves represent SKA1-LOW Deep, SKA1-LOW and LOFAR, respectively.}
\label{obs}
\end{figure*}

Finally, Fig.~\ref{obs} shows the conditional probability, $P_\mathrm{obs}$, that an exoplanet can be detected as a function of the source distance, given that the host star enters the Einstein radius of a lens star.
Red, blue and green curves represent SKA1-LOW Deep, SKA1-LOW and LOFAR, respectively. In upper and lower panels, the scaling law of $P_\mathrm{rad} \propto a^{-2}$ and $a^{-4/3}$ are assumed, respectively. 
In left and right panels, the lens mass is set to $0.1~M_{\odot}$ and $1.0~M_{\odot}$, respectively. For simplicity, the lens distance is set to the half of the source distance. The semi-major axis is chosen randomly according to a probability distribution function derived from the observed hot-Jupiter population ($0.00263~\mathrm{AU} - 0.1~\mathrm{AU}$) \footnote{ \href{http://exoplanet.eu}{\textit{http://exoplanet.eu}}}, which is close to Gaussian distribution with mean of $0.05~\mathrm{AU}$ and standard deviation of $0.02~\mathrm{AU}$. The values of $u_\mathrm{min}$, $\phi$, $i$ and $\Omega$ are randomly chosen from uniform probability distributions. Other parameters are set as $M_{\star} = 1.0~M_{\odot}$ and $V = 200~\mathrm{km/s}$. 

It is seen that the probability does not depend largely on the lens mass. This is because the magnification is determined by $u$, which is the relative angle between the source and lens normalized by the Einstein radius. Therefore, although the probability that a source is microlensed increases as the lens mass increases, the conditional probability of the current interest is not affected. On the other hand, the observability strongly depends on the scaling law. In fact, the difference in the radio emissivity at $0.05~\mathrm{AU}$ is a factor of 20 between the two scaling laws, noting that it is normalized by Jupiter at $5~\mathrm{AU}$.

To estimate the event rate of microlensing of hot Jupiters, we focus on exoplanets located at the bulge region, which will dominate microlensing events toward the Galactic Center. Then, we take $P_\mathrm{obs} = 3 \times 10^{-3}$ and $3 \times 10^{-5}$ as typical probabilities for exoplanets with scaling law of $a^{-2}$ and $a^{-4/3}$, respectively, considering the SKA1-LOW Deep observation. In this case, we have an event rate of $R_\mathrm{lens} \times P_\mathrm{HJ} \times P_\mathrm{obs} \times S \approx 0.3~\mathrm{year}^{-1}$ and $0.003~\mathrm{year}^{-1}$ for $a^{-2}$ and $a^{-4/3}$ scalings, respectively. In case of SKA1-LOW and LOFAR, the expected event rates are lower than the above by about a factor of 10 and 300, respectively.
    
\section{Summary and Discussion}
\label{s5}

In this paper, we studied microlensing of a exoplanetary system as a source, not as a lensing object. This enhances the detectability of the radio emission which allows us to probe the activity of the planet's magnetosphere \citep{2011A&A...531A..29H,zarka2015}. Due to the orbital motion of the exoplanet around the host star, if the semi-major axis is relatively small ($\lesssim 0.1~\mathrm{AU}$), the magnification curve has a wavy feature which reflects the orbital parameters and masses of the host and lens stars. This is a unique feature which can be used to identify microlensing event of an exoplanet.

Auroral radio emission from exoplanets, especially hot Jupiters, can be much brighter than the host star in low-frequency radio band and can extend to $\gtrsim 50~\mathrm{MHz}$ for an exoplanet with relatively strong surface magnetic fields ($\gtrsim 20~\mathrm{Gauss}$). Thus, auroral radio emission of exoplanets is an intriguing target of the next-generation radio telescopes such as the SKA-LOW, and microlensing can enhance the observability.

We estimated the event rate of microlensing of exoplanets toward Galactic Center region expected with monitoring observations by the SKA1-LOW and LOFAR. It was found that the expected event rate is as large as $0.03~\mathrm{year}^{-1}$ and $0.3~\mathrm{year}^{-1}$ by SKA1-LOW observation with $4~\mathrm{MHz} \times 1~\mathrm{hours}$ and $40~\mathrm{MHz} \times 10~\mathrm{hours}$ integrations, respectively. Although the event rate is not high even for the SKA1-LOW, magnification by microlensing gives us a chance to observe distant exoplanets.

Our discussion on observability relied largely on the scaling law of radio emissivity, which is highly uncertain at this time. In addition, while we estimated the auroral radio power by simply scaling the Jupiter's value, it would also be affected by the properties of the stellar wind, and different types of the host stars may have a different normalization.
A different scaling law and/or normalization can lead to substantially different results and, conversely, statistics of radio emissivity of exoplanets can be probed from future observations.

It is well known that auroral radio emission of Jupiter and Saturn are highly anisotropic \citep{Lamy2008}.
Therefore, if the radio emission of hot Jupiters are also anisotropic, the event rate will be reduced by a factor of the solid angle of the emission divided by $4 \pi$.

In this paper, we considered radio emission from the magnetosphere of exoplanets. Another possible radio emission involving exoplanets is that from the star-planet interaction. Most M dwarfs have short-period planets and M dwarf-planet systems can be promising radio sources \citep{vedantham2020}. While microlensing can amplify these systems and enhance the observability, the characteristic wavy feature is not likely to be seen when the dominant radio emission comes from the surface of the host star.

It should be noted that the shape of the light curve of a microlensed hot Jupiter does not necessarily coincide with that of the magnification curve, because the radio emission is expected to have an intrinsic variability. In fact, Jupiter's auroral radio emission varies with its spin period. Nevertheless, if the timescale of intrinsic variability is shorter than the orbital period and crossing time, the wavy feature could still be observed in the light curve. Observation of the magnification curve is very important because the wavy feature contains much information on the system such as the masses of the host and lens stars and the orbital parameters.

In this paper, we considered only exoplanets as radio sources toward the Galactic Center. Actually, the observation of radio emission from exoplanets could be confusion-limited due to a finite angular resolution of the radio telescopes. To identify exoplanets, the circular polarization is a good measure because most of other sources are expected to be unpolarized. Also, as was shown in our manuscript, micro-lensed exoplanets can be distinguished by the unique feature of the magnification curve if a significant fraction of the light curve, not just the peak, could be observed. Furthermore, if optical data is available, the cross-matching between radio (circular polarization) and optical sources is an effective way to reduce the confusion \citep{Callingham2019}.

The synergy between the radio and optical/infrared observations is a promising way to not only identify micro-lensed exoplanets but constrain their orbital parameters. In fact, as we saw in Fig.~\ref{fig_p_ex}, the magnification curve is different between the host star and the planet, and they will provide us complimentary information. A quantitative discussion on parameter estimation, considering OGLE, MOA and WFIRST as the optical counterparts of the SKA, will be given elsewhere.

\section*{Acknowledgements}

KT is partially supported by Grand-in-Aid from the Ministry of Education, Culture, Sports, and Science and Technology (MEXT) of Japan, No. 15H05896, 16H05999 and 17H01110, and Bilateral Joint Research Projects of JSPS.
YF is supported by Grand-in-Aid from MEXT of Japan, No. 18K13601. 


\bibliographystyle{mnras}
\bibliography{ref}

\label{lastpage}
\end{document}